\documentclass[aps,preprintnumbers,twocolumn,superscriptaddress,showpacs,prl,floatfix]{revtex4}

\usepackage{graphicx}

\begin{document}

\preprint{TTK-10-19, WUB/10-05}

\title{Dominant QCD Backgrounds in Higgs Boson Analyses at the LHC: \\
       A Study of {\boldmath $pp \rightarrow t \bar t$} + 2 jets at
       Next-To-Leading Order}

\author{G. Bevilacqua}
\affiliation{Institute of Nuclear Physics, NCSR Demokritos, GR-15310
             Athens, Greece}
\author{M. Czakon}
\affiliation{Institut f\"ur Theoretische Teilchenphysik und
  Kosmologie, RWTH Aachen University, D-52056 Aachen, Germany}
\author{C. G. Papadopoulos}
\affiliation{Institute of Nuclear Physics, NCSR Demokritos, GR-15310
             Athens, Greece}
\author{M. Worek}
\affiliation{Fachbereich C, Bergische Universit\"at Wuppertal, D-42097
             Wuppertal, Germany}

\begin{abstract}

We report the results of a next-to-leading order simulation of top
quark pair production in association with two jets. With our inclusive
cuts, we show that the corrections with respect to leading order are
negative and small, reaching 11\%. The error obtained by scale variation is of
the same order. Additionally, we reproduce
the result of a previous study of top quark pair production in
association with a single jet.

\end{abstract}


\maketitle

Exploiting the relatively clean experimental signal and excellent
theoretical understanding of the $W^+W^-$ Higgs boson decay channel,
the Tevatron collaborations continue excluding Higgs boson masses,
$M_H$, in the vicinity of twice the $W$ boson mass with ever growing
confidence levels \cite{Aaltonen:2010yv}. As a consequence, 
studies assuming a lighter
Higgs boson are more timely than ever. Clearly, it is  necessary to
devise suitable scenarios that would allow for discovery and
measurement of the basic properties of the scalar at the Large Hadron
Collider (LHC). Among the many available scenarios, one is especially
disputed. It has been argued that if $M_H < 135$ GeV, the production
and decay chain $pp \rightarrow t\bar tH^* \rightarrow t\bar t b\bar
b$ should provide an excellent opportunity owing to the large top and
bottom quark Yukawa couplings. The question of discrimination between
the large backgrounds and signal was expected to be solved by the
sharp Breit-Wigner peak of the Higgs boson in the invariant mass
distribution of
the b quark jets. Unfortunately, application of realistic selection
cuts and acceptances demonstrated a smearing of the resonance far
beyond what would be expected from initial state radiation 
\cite{Ball:2007zza,Aad:2009wy}. At this point it seems, therefore, that 
a very precise  knowledge of the
backgrounds is necessary, if the channel is to be of any
usefulness \footnote{We point, however, to recent ideas \cite{Plehn:2009rk}, 
which may change the situation.}.

\begin{figure}[t]
{\includegraphics[width=.40\textwidth]{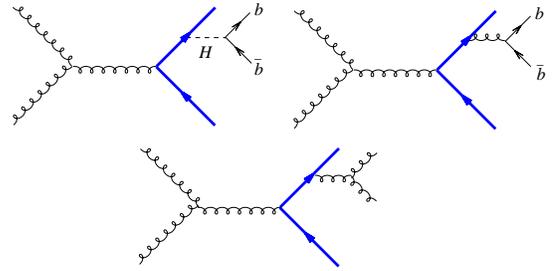}}
\vspace*{-1em}
\caption{Example diagrams contributing to the signal $t\bar tH^*
  \rightarrow t\bar tb\bar b$, and the irreducible background with the
same final state, as well as the reducible background with two
jets. Thick blue lines correspond to top quarks, the wiggly to gluons
as usual.}
\label{fig:fd}
\end{figure}

A close scrutiny of the backgrounds shows \cite{Ball:2007zza,Aad:2009wy}
that the most relevant are the direct production of the final state $t\bar tb\bar b$
(irreducible background) and the production of a top quark pair in association
with two jets, $t\bar t jj$ (reducible background). The latter needs
to be taken into account due to the finite efficiency in identifying
b-quarks in jets (b-tagging). We depict example diagrams contributing
to the signal and the two backgrounds in Fig.~\ref{fig:fd}. It is
crucial that although $t\bar tjj$ has a cross section, which is
larger than that of $t\bar tb\bar b$ by about two orders of
magnitude, in the actual setup of \cite{Aad:2009wy}, both backgrounds turn out to give
very similar contributions. Recent studies 
\cite{Bredenstein:2009aj,Bevilacqua:2009zn,Bredenstein:2010rs} have demonstrated that
the next-to-leading order QCD corrections to $t\bar tb \bar b$ are
very large, undermining even further the feasibility of actual analyses
in the channel at hand. A proposed solution \cite{Bredenstein:2010rs} involves the use
of a dynamical renormalization/factorization scale in simulations and,
more importantly, the imposition of a jet veto on the third jet (upper
bound on the allowed transverse momentum, $p_T$). The purpose of this
letter is to determine, whether large corrections affect the $t\bar
tjj$ channel.

%------------------------------------------------------------------------------%
%------------------------------------------------------------------------------%

In principle, the concept of a cross section of a process involving
massless partons at leading order (LO), as is the case of $t\bar tjj$, is
undefined due to potential soft and collinear divergences and requires
the specification of separation cuts. This complicates somewhat our
problem, since any conclusions we will draw will pertain to the
particular setup we will have used. Therefore, in order to retain some
level of generality, we will not attempt to reproduce the exact
conditions of the Higgs boson analysis of \cite{Aad:2009wy}. On the
 contrary, we will mimic some of the general assumptions, but impose 
more inclusive cuts.

To be specific we consider proton-proton collisions at the LHC with a
center of mass energy of $\sqrt{s}=14$ TeV. We set the mass of the top
quark to be $m_t=172.6$ GeV  and leave it on-shell with unrestricted
kinematics. The jets are defined by at most  two partons using the
$k_T$ algorithm of \cite{Catani:1992zp,Catani:1993hr} 
with a separation $\Delta R=0.8$, where $\Delta
R=\sqrt{(y_1-y_2)^2+(\phi_1-\phi_2)^2}$,
$y_i=1/2\ln(E_i-p_{i,z})/(E_i+p_{i,z})$ being the rapidity and
$\phi_i$ the azimuthal angle of parton $i$. Moreover, the recombination
is only performed if both partons satisfy $|y_i|<5$ (approximate
detector bounds). We note that the $k_T$ algorithm specifies not only
which partons are combined into jets, but also the momentum of the
resulting jets. In our simulation, we assume that the four-momenta of
the partons are added. We will comment on the importance of this
point, when we compare with the $t\bar tj$ calculation of 
\cite{Dittmaier:2007wz,Dittmaier:2008uj}. We
further assume that the jets are separated by $\Delta R=1$ and have
$|y_{\rm{jet}}| < 4.5$. Their transverse momentum is required to be
larger than 50 GeV. It is mostly in $\Delta R$ and $p_{T,min}$ that
the present setup is different from that used in the case of $t\bar
tb\bar b$ \cite{Bredenstein:2009aj,Bevilacqua:2009zn}. 
Here, we use a higher $p_T$ cut similarly to the $t\bar
tj$ case of \cite{Dittmaier:2008uj}. The additional separation in $\Delta R$ is only used
to demonstrate the flexibility of our tools, but is believed to bear
no impact on the final conclusions. Notice, that our jets are allowed
to contain b quarks, and therefore also the final state $t\bar tb\bar
b$. This is irrelevant, since we know that this contribution is tiny
and can be simply subtracted from our results (the summation is
incoherent).  Finally, we note that the third jet, which stems from
real radiation, is not restricted. Nevertheless, we will also study
the impact of a jet veto with $p_T=50$ GeV.

The production of jets in hadronic collisions can be decomposed into
many processes at the parton level. In Tab.~\ref{tab:lo}, we summarize
the LO contributions of different subclasses. These were obtained with
the kinematics specified above using the CTEQ6L1 LO parton
distribution functions \cite{Pumplin:2002vw,Stump:2003yu} 
and the LO running of the strong
coupling constant up to the scale (common for renormalization and
factorization) $\mu_0=m_t$. There are two interesting conclusions to
be drawn here. First, it comes as a surprise that the mixed channel
$qg$ is more important than $gg$. The proportions between the two
depend, however, on the $p_T$ cut. If $p_{T,min}=20$ GeV, the
situation is reversed. The reason is that the lower the cut, the
larger the soft gluon enhancement. If there are more gluons
in the final state the enhancement will be higher, and therefore at
some point $gg \rightarrow t\bar tgg$ must be larger than $qg
\rightarrow t\bar tqg$. The second, more important, point is that the
channels related to the $t\bar tb\bar b$ final states, in particular
$gg \rightarrow t\bar tq \bar q$ are almost negligible compared to the
two dominant (we referred to this before). This implies that whatever
the result on the size of the corrections in the $t\bar tb\bar b$
case, $t\bar tjj$ requires a separate study.

%------------------------------------------------------------------------------%
\begin{table}
\vskip .4 cm
\begin{tabular}{|c|c|c|}
\hline
\textsc{Process} & $\sigma^{\rm {LO}}$ [pb] & \textsc{Contribution} \\
\hline
\hline
$pp\rightarrow t\bar{t}jj$   &  120.17(8)  &  100 \% \\
\hline
\hline
$qg\rightarrow t\bar{t} qg$   & 56.59(5)  &  47.1  \% \\
\hline
$gg\rightarrow t\bar{t} gg$   & 52.70(6)   & 43.8 \% \\
\hline
$qq'\rightarrow t\bar{t} qq'$, ~$q\bar{q}\rightarrow t\bar{t} q'\bar{q}'$
 & 7.475(8)          & 6.2 \% \\
\hline 
$gg\rightarrow t\bar{t} q\bar{q}$ &  1.981(3)   &  1.6 \% \\
\hline
$q\bar{q}\rightarrow t\bar{t} gg$ & 1.429(1)   &   1.2 \% \\
\hline
\end{tabular} 
\caption{The LO cross section for $pp\rightarrow t\bar{t}jj$ production 
at the LHC. The individual contributions of the various partonic channels are
also presented  separately. Both $q$ and $q'$ span
all quarks and anti-quarks.}
\label{tab:lo}
\end{table}
%------------------------------------------------------------------------------%

Before we give our results for the next-to-leading order (NLO)
corrections, we are compelled to present the computational framework
used for the simulations. Similarly to our previous 
publication \cite{Bevilacqua:2009zn}, we 
have used the \textsc{Helac-Phegas} framework 
\cite{Kanaki:2000ey,Papadopoulos:2000tt,Cafarella:2007pc} and
in particular \textsc{Helac-1L} \cite{vanHameren:2009dr} for the evaluation 
of the virtual corrections. This tool uses the OPP 
method \cite{Ossola:2006us} 
and \textsc{CutTools} \cite{Ossola:2007ax,Ossola:2008xq,Draggiotis:2009yb} 
for the reduction of tensor integrals, as well as \textsc{OneLOop} for 
numerical values of scalar integrals. The practical techniques involve
re-weighting of events and sampling over polarization and color, for
details see \cite{Bevilacqua:2009zn}. The real radiation corrections
were evaluated with \textsc{Helac-Dipoles} \cite{Czakon:2009ss}, 
which is an implementation of the Catani-Seymour subtraction 
formalism \cite{Catani:1996vz,Catani:2002hc}. In order to check our
results, we have explored the independence of the results on the
unphysical cutoff in the dipole subtraction phase space, 
see \cite{Czakon:2009ss} and references therein for details. We have also
verified the cancellation of divergences between the real and virtual
corrections. Finally, the numerical precision of the latter was
assured by using gauge invariance tests and use of 
quadruple precision. We note that the only new virtual amplitudes are
those involving a top quark pair and four gluons. These were presented
for the first time in \cite{vanHameren:2009dr} and, due to their notorious
complexity, still await an independent check by other groups.

For the evaluation of the NLO corrections, we have used the CTEQ6M
parton distribution functions with NLO running of the strong coupling
constant. At the central scale $\mu_0=m_t$, we obtain
\[
\sigma^{\rm{NLO}}_{pp\rightarrow t\bar tjj+X} = (106.94 \pm 0.17) ~\rm{pb} \;,
\]
where the error comes from Monte Carlo integration. 
Compared to the LO  result from Tab.~\ref{tab:lo}, this
represents a negative shift of 11\%, and allows us to conclude that
the corrections to this process are small.

\begin{figure}[t]
{\includegraphics[width=.40\textwidth]{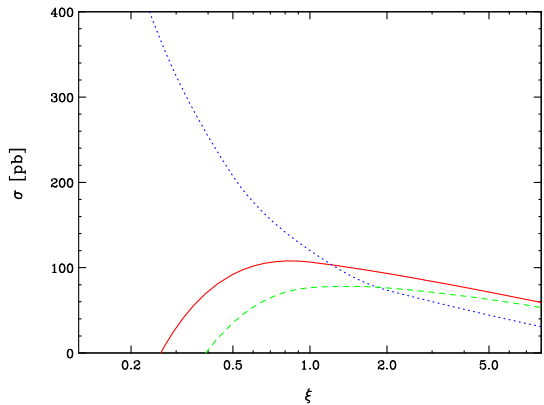}}
\vspace*{-1em}
\caption{Scale dependence of the total cross section for
  $pp\rightarrow t\bar{t} jj  + X$ at the LHC   with $\mu_R=\mu_F=\xi
  \cdot \mu_0$ where $\mu_0=m_t$.  The blue dotted curve corresponds
  to  the LO, the red solid to the NLO  result whereas the green
  dashed to the NLO result with a jet veto of 50 GeV.}
\label{fig:scales}
\end{figure}

The scale dependence of the corrections is illustrated in
Fig.~\ref{fig:scales}. At first, we observe a dramatic
reduction of the scale uncertainty while going from LO to NLO. Varying
the scale up and down by a factor 2 changes the cross section by
+72\% and -39\% in the LO case, while in the NLO case we have
obtained a variation of -13\% and -12\%. Second, the central
scale that we have chosen is very close to the point of minimal
corrections and slightly above the point of maximum of the NLO cross
section. Indeed, both $\mu=1/2\mu_0$ and $\mu=2\mu_0$ give smaller
values.

Taking into account the above dependence on the scale choice, it is to
be expected that adding a jet veto will only worsen the result. In
order to make this more transparent, it is best to consider the
difference between 
the cross section without and with the jet veto. This difference is
given by a LO  calculation, since it requires the existence of
three separated jets with a lower cut on their respective transverse
momenta. Clearly, not only will it be negative, but it will also have
to grow more negative with diminishing scale (since this is almost
entirely governed by the behavior of the strong coupling constant
alone). At this point it is difficult to decide whether a jet veto
will or will not be necessary for the complete Higgs boson analysis. What we
can give is the lower bound on the corrections, assuming
that a jet veto of less than $p_T=50$ GeV is likely to endanger the
stability of the perturbative expansion. The total cross section with
a jet veto of 50 GeV is
\[
\sigma^{\rm{NLO}}_{pp\rightarrow t\bar tjj+X}(p_{T,X} < 50 ~ \rm{GeV})
= (76.58 \pm 0.17) ~\rm{pb} \; ,
\]
with a scale variation of -54\% and -0.3\%, see
Fig.~\ref{fig:scales}. The plots show that choosing a
higher scale in the case of the jet veto, would lead to a result with
virtually no scale dependence. This should be considered as severely
underestimating the error.

%------------------------------------------------------------------------------%
%------------------------------------------------------------------------------%

\begin{figure}[ht]
{\includegraphics[width=.40\textwidth]{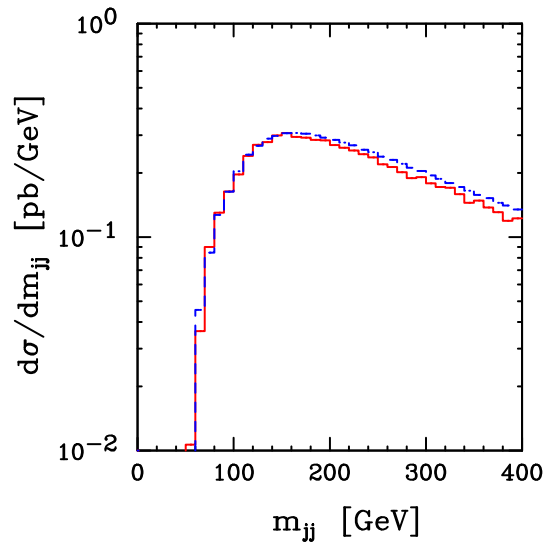}}
\vspace*{-1em}
\caption{
Distribution of the invariant mass $m_{jj}$ of the first and the second
hardest jet for $pp\rightarrow t\bar{t} jj +X$ at the LHC.  The red solid 
line refers to the NLO result while the blue dotted  line to the LO  one.}
\label{fig:mjj}
\end{figure}

While the size of the corrections to the total cross section is
certainly interesting, it is crucial to study the corrections to the
distributions. The most important for us is the invariant mass of the
two tagging (highest $p_T$) jets, since this is the observable
entering Higgs boson studies. We plot the LO and NLO results in
Fig.~\ref{fig:mjj}. While we notice a long tail, we keep the
dependence only in a modest range up to 400 GeV due to our
phenomenological motivation. The distribution starts above about 45
GeV due to the $\Delta R$ and $p_T$ cuts, and shows tiny corrections
up to at least 200 GeV, which means that the size of the corrections
to the cross section is transmitted to the most relevant distribution.

%------------------------------------------------------------------------------%
%------------------------------------------------------------------------------%

%------------------------------------------------------------------------------%
\begin{figure}[ht]
{\includegraphics[width=.40\textwidth]{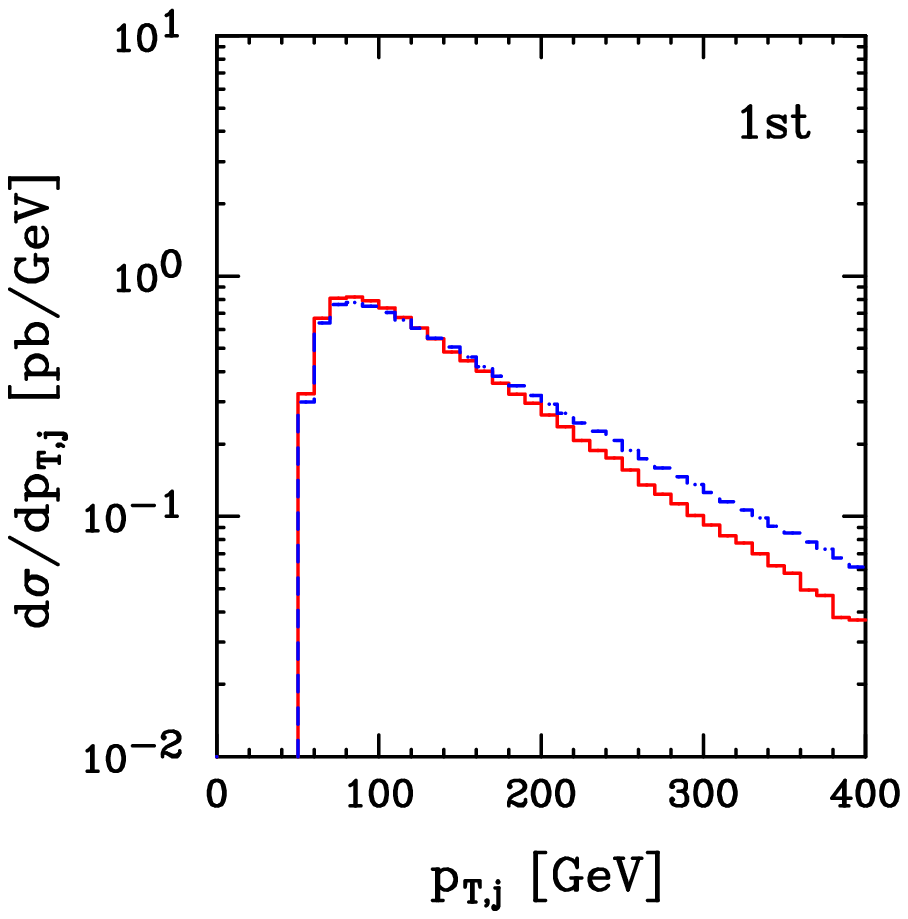}}
\vspace*{-1em}
\caption{Distribution in the transverse momentum $p_{T_{j}}$  of the  
1st hardest jet for $pp\rightarrow t\bar{t} jj +X$ at the LHC.  The 
red solid line refers to the NLO result while the blue dotted  line 
to the LO  one.}
\label{fig:pTj1}
\end{figure}
%------------------------------------------------------------------------------%

%------------------------------------------------------------------------------%
\begin{figure}[ht]
{\includegraphics[width=.40\textwidth]{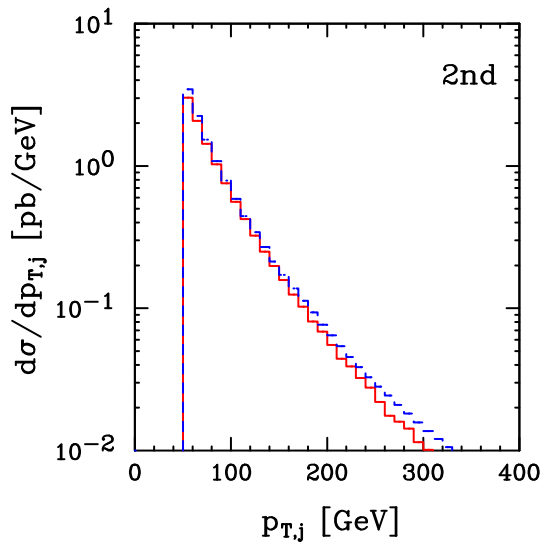}}
\vspace*{-1em}
\caption{Distribution in the transverse momentum $p_{T_{j}}$  of the  
2nd hardest jet for $pp\rightarrow t\bar{t} jj +X$ at the LHC.  The 
red solid line refers to the NLO result while the blue dotted  line 
to the LO  one.}
\label{fig:pTj2}
\end{figure}
%------------------------------------------------------------------------------%

Of course, there are observables showing much larger effects. The
classic example is the transverse jet momentum distribution at high
$p_T$. We illustrate the phenomenon in Figs.~\ref{fig:pTj1} and
\ref{fig:pTj2}, which demonstrate the strongly altered shapes in the
cases of the hardest and second hardest jets. It is well known that
this kind of corrections can only be correctly described by higher
order calculations. On the other hand, the behavior at low $p_T$ is
certainly further altered by soft-collinear emissions, which are best
simulated by parton showers. With our lower cut of $p_{T,min}=50$ GeV,
we expect to be mostly in the safe range, where fixed order
perturbation theory does not break down.

While the above comments conclude our analysis of  $t\bar tjj$, we
would like to make a few statements about the process with only one
jet, $t\bar tj$. While preparing our calculation, we have made a
comparison with the results of \cite{Dittmaier:2008uj}. Using 
exactly the same setup as
that work, we were able to obtain the following value for the cross
section
\[
\sigma^{\rm{NLO}}_{pp\rightarrow t\bar tj+X} = (376.6 \pm 0.6) ~
\rm{pb} \; ,
\]
demonstrating perfect agreement
(\cite{Dittmaier:2008uj} quotes $(376.2 \pm 0.6)$ pb). We note that besides a
different value of the top quark mass ($m_t = 174$ GeV) and $\Delta
R=1$ in the jet algorithm, the authors used a $k_T$ algorithm, where
the momenta of the partons being combined into a jet are not simply
added. Instead, massless jets are constructed using transverse momenta
(which are trivially added) and rapidities. The version that we use
would lead to a lower value of the cross section, since the transverse
momentum of the sum of momenta is never larger than the sum of the
transverse momenta, and the only relevant cut is applied to the jet
$p_T$. Our value would instead be $(372.2 \pm 0.6)$~pb. On the other
hand, with all parameters and cuts as given in the introduction the
final result is $(376.1 \pm  0.7)$~pb.

\vspace{0.2cm}
Let us conclude by pointing out that while we demonstrated that the
corrections in $t\bar tjj$ are small for one setup, comparison to the
behavior of the corrections for different setups in the $t\bar tj$
case \cite{Dittmaier:2008uj} provides a viable
 argument that our conclusions will remain true for other input
parameters. Nevertheless, we are planning to present a much wider
study involving in particular a variation of the center of mass
energy, cone size in the jet algorithm, transverse momentum cuts and
jet vetoes.

\vspace{0.2cm}
\begin{acknowledgments}
The simulations presented in this 
work have been performed at the DESY
Zeuthen Grid Engine computer cluster.  The work of M.C. was supported
by the Heisenberg Programme of the Deutsche
Forschungsgemeinschaft. The authors were funded in part by  the RTN
European Programme MRTN-CT-2006-035505 HEPTOOLS - Tools and Precision
Calculations for  Physics Discoveries at Colliders. M.W. was
additionally supported by the Initiative and Networking Fund of the
Helmholtz Association, contract HA-101 (”Physics at the Terascale”).
\end{acknowledgments}

\end{document}